# A Statistical Approach to Vehicular Traffic

Jan Freund[*]

*Humboldt–Universität zu Berlin, Institut für Physik,*

*Unter den Linden 6, D–10099 Berlin, Germany*

Thorsten Pöschel[†]

*Arbeitsgruppe Nichtlineare Dynamik, Universität Potsdam,*

*Am Neuen Palais, D–14415 Potsdam, Germany, and*

*The James Franck Institute, The University of Chicago,*

*5640 South Ellis Avenue, Chicago, Illinois*

(May 21, 1995)

## Abstract

A two–dimensional cellular automaton is introduced to model the flow and jamming of vehicular traffic in cities. Each site of the automaton represents a crossing where a finite number of cars can wait approaching the crossing from each of the four directions. The flow of cars obeys realistic traffic rules. We investigate the dependence of the average velocity of cars on the global traffic density. At a critical threshold for the density the average velocity reduces drastically caused by jamming. For the low density regime we provide analytical results which agree with the numerical results.
PACS numbers:

Typeset using REVTEX

[*]E–mail: janf@itp02.physik.hu–berlin.de

[†]E–mail: thorsten@hlrsun.hlrz.kfa–juelich.de, http://summa.physik.hu-berlin.de:80/~thorsten/

# I. INTRODUCTION

Vehicular traffic on highways as well as in cities tends to suffer from a jamming transition when the global traffic density exceeds a critical threshold value. The phenomena related to traffic jams have attracted the attention of engineers and physicists since many years and there exists a large variety of experimental observations, see [1–4] and references therein. The formation of traffic jams has been investigated by many authors using various methods. As early as 1955 Lighthill and Whitham [5,6] described the spontaneous formation of regions of increased car concentrations and shock waves using their approach of kinematic waves, i.e. wave motion where the spatial value of the flow is a function of the spatial concentration distribution. Prigogine and Herman [7,8] investigated traffic jams on highways by performing a stability analysis of hydrodynamic equations. One–dimensional cellular automata models [9] for the simulation of traffic flow have been proposed e.g. in [10–12]. This type of models has been studied intensively using massively parallel computers [13–17]. Schadschneider and Schreckenberg [18] solved a one–dimensional cellular automaton model analytically in the mean field approximation, Csahók and Vicsek [19] investigated their automaton in the presence of quenched noise. Nonlinear wave descriptions have been proposed in [20–22]. Ben–Naim et. al [23] apply a ballistic aggregation process to model the kinetics of clustering in one–dimensional traffic flow. Hydrodynamic approaches have been investigated by various authors, e.g. [24,25]. These models are based on certain assumptions concerning the "hydrodynamic" properties of traffic flows such as the velocity versus density relation [26,27], and the viscous terms [22].

One–dimensional traffic has been studied extensively and the developed models are sufficiently sophisticated to reproduce experimental observations, i.e. the fundamental diagram (throughput versus flow density) of one lane roads. Some of the authors (e.g. [28]) claim that the size distribution of the jams obeys a power law, but very large scale computations have shown that the lifetimes of the jams and hence their sizes reveal a characteristic cut–off which seems to be no finite size effect of the simulation [14]. There are several models describing traffic flow in more than one dimension. The analysis of traffic on a network with multiple sources and sinks can be found e.g. in [29–32]. Nagatani [33] describes a cellular automaton representing a two–lane roadway.

An interesting new approach was currently proposed by Bando et al. [34] who express the rules which govern the behavior of the cars (i.e. acceleration and deceleration) by a dynamic equation for each car $\ddot{x}_n = \alpha \{V(x_{n+1} - x_n) - \dot{x}_n\}$. The interaction of neighboring cars is expressed in terms of the function $V$ of the distance between the cars. Using this approach the authors connect in some respect the model of the cellular automata with molecular dynamics.

Two–dimensional cellular automata designed to simulate traffic in a city have been investigated by some authors. Biham et al. [35] proposed a three state cellular automaton model where at a given time step each site can be occupied by a car moving from South to North, by a car moving from West to East, or the site may be empty. The time behavior is ruled by synchronous traffic lights at each site, allowing alternatively for vertical or horizontal traffic. Obviously there are situations when the traffic jams, i.e. when the traffic light allows for driving but the next site is occupied. The model given in [35] and particularly the spatial correlations in the jamming phase have been numerically investigated by Tadaki and Kikuchi [36]. A similar model but with "faulty" traffic lights was studied in [37]. Nagatani [38] investigated the spreading of a jam which is induced by an accident using an extremely simple automaton rule. Numerically he finds scaling laws for the size of the spreading jam as a function of time elapsed since the occurrence of the accident which cannot be derived analytically so far. Fukui et al. [39] numerically investigated the evolution of ensemble averages of the jamming process for the simple automaton model described in [35]. For the same model [35] Chau et al. [40] recently gave an analytic upper bound (depending on the dimension) for the critical car density $\eta_{cr}$, i.e. when the system transits from the "moving phase" into the "jamming phase". In two dimensions they found $\eta_{cr} \leq 11/12$. Obviously this upper bound is not close to the value which was observed in numerical simulations. In the case of Nagatani and Seno [41] there is a set of parallel one–way streets, all oriented in the same direction ($x$–direction), which is intersected by a single perpendicular one–way street in $y$–direction.

For fixed car density on the perpendicular street $\rho_y$ they find that the flux of cars $J_x$ rises linearly with the car density $\rho_x$ until a characteristic critical threshold $\rho_x^c$ is reached. When further increasing the density $\rho_x$ the flux $J_x$ approximately remains constant while the average velocity of the cars $\langle v_x \rangle$ drops. This rather sharp change in the overall behavior is due to the formation of traffic jams where the flow is irregular and discontinuous. When further increasing the density $\rho_x$ the discontinuous character of the flow disappears at a second threshold $\rho_x^C$ and the flux $J_x$ declines linearly. This behavior corresponds to the symmetry of the car density in $y$–direction and the spaces between the cars $1 - \rho_y$. Indeed, one finds that the critical densities $\rho_x^c$ and $\rho_x^C$ are almost exactly symmetric, e.g. for $\rho_y = 0.3$ one finds $\rho_x^c = 0.31$ and $\rho_x^C = 0.69$.

Cuesta et al. [42,43] have been the first who reported on simulations of a two–dimensional automaton where the cars are allowed to change their direction. In their model each crossing can be occupied by one car or it can be empty. Each car is assigned a preferred direction; the parameter $w_i(r)$ is the probability that the $i$-th car at the position $r$ moves in the next time step horizontally and $1 - w_i(r)$ is the probability to move in vertical direction. Horizontal motion is allowed at even time steps, vertical motion at odd times. There have been defined two variants, A: there are only one–way streets directed from South to North and from East to West. Half of the cars is given the trend $w_i(r) = \gamma$ and the other half is given $w_i(r) = 1 - \gamma$. In the second variant B there are one–way streets pointing alternatively South $\to$ North and North $\to$ South, and West $\to$ East and East $\to$ West, respectively. The trends of the cars which are subdivided into four equal groups, point into one of the four possible directions.

We should mention that several authors claim that at least the one–dimensional traffic flow problem is closely related to sand flowing in pipes, e.g. [44–48].

In the present paper we are concerned with a two–dimensional cellular automaton designed to simulate traffic in a city. As a special feature of our model we emphasize the fact that we use rather realistic traffic rules which will cause a slowing down of the average traffic velocity and finally will lead to the collapse.

## II. THE MODEL

In our model the city is represented by a set of $L$ streets in horizontal direction crossing $L$ streets in vertical direction. Cars are allowed to move in both directions, i.e. we have no restrictions to one–way streets. The crossings define a two–dimensional cellular automaton, each of the $L^2$ crossings is represented by a site of the automaton. We assume periodic boundary conditions in both directions. Fig. 1 shows a schematic plot of an automaton site (a crossing). At each crossing there are four queues of maximum length $Q$ filled by cars coming from one of the four directions, representing the finite space where cars can move freely between crossings in realistic urban traffic systems. Hence, at each crossing there are allowed at most 4 $Q$ cars. Each of the cars has a desired direction, i.e. to the right, to the left or straight on, which has to be chosen according to a certain rule. We will discuss this point in detail below. Within each time step the first of the cars in each queue can go to the next crossing provided that it does not have to give way to one of the other cars at the same site. In case it has to give way to another car it will not move during this time step. In our model we assume the simple and realistic rule that a car has to give way in the case that another car occupies the same crossing at its right hand side. Moreover a car has to stop if it intends to turn left and there is a car at the opposite side of the crossing which goes straight on or turns to the right. If all of the four top positions of a crossing are occupied by four cars one of the cars will be selected randomly and will then be allowed to move in the current time step. Insofar we have chosen realistic rules which are current law in many countries.

Each of the $N$ cars in the system starts at a randomly selected crossing, its desired direction (left, straight on, right) will be determined according to some rule (discussed below). Within each time step a car maximally moves a distance of one lattice cell, i.e. it tries to transit to one of the next neighbors of the current site provided the following three conditions hold:

1. It does not have to give way.

2. It is placed at the top position of its queue.

3. There are less than $Q$ cars standing in the destination queue.

Otherwise it rests. Obviously, if there is only one car in town it never has to give way nor will it find another car in front of itself. Therefore the average velocity[1] $\langle v \rangle = 1$ in units of lattice cells per time steps.

As mentioned above we have to discuss the rule which determines the desired direction within each time step. There are at least three reasonable rules:

1. The desired direction for each car is selected at random within each time step not regarding whether the car moved in the previous time step or stopped. This rule is rather artificial since provided a jam occurs it can dissolve due to the fact that the cars might give up their previously desired direction when they cannot move.

2. Each car is assigned a destination site on the lattice. Once a car reached its destination it will be assigned a new randomly chosen destination. This rule is problematical since one has to define the detailed path a car follows to reach its destination. A natural choice would be to chose a path so that the car has to change its direction only once on its way between its starting position and its destination. In this case, however, the overall behavior of the system would depend on the size of the lattice $L$ since one can check that the probability to be allowed to move straight on is higher than to be allowed to turn. Cars which intend to turn either left or right (on the average) have to give way more frequently than those which go straight on. The larger the system the smaller becomes the probability per time step to turn. For the case $L \to \infty$ the probability for a car in a given time step to turn approaches zero. Hence this rule does not allow for size–independent results, and the results of a simulation using this rule apply only for one fixed lattice size $L$.

---

[1] Here and in the following we describe the ensemble average of a value $x$ by the symbol $\langle x \rangle$, whereas the time average by $\overline{x}$.

3. Each car is assigned a desired direction during the initialization. But now, only after moving one step to the aimed destination a new desired direction will be chosen at random, i.e. the primarily chosen desired direction is maintained by each car until all of the three conditions hold which allow the car to move. This rule has several advantages compared with the other rules. First we avoid the size dependent behavior of the previous rule, and second we do not have long range correlations between the lattice sites for the case of low car density, which would be caused by the second rule. Hence the behavior of the system can be described by a Markov process, which will be the basic assumption of our analytical calculation in section IV. An analytical description which takes into account the long range correlation between the sites coming from the rules which determine the path of the cars seems to be very difficult, except perhaps for the (trivial) case $L \to \infty$, when the probability of the cars to turn, and hence the traffic rules connected with the turn, can be neglected. In the following we always refer to the third rule.

## III. NUMERICAL RESULTS

In this section we present the results found by numerical simulations of a cellular automaton which represents the traffic rules described above.

The two–dimensional automaton consists of $L \times L$ sites on a rectangular lattice. To eliminate boundary induced effects we have chosen periodic boundary conditions in both directions, i.e. we assume the topology of a torus. We simulated systems of size $L = 20$, 30, 50, 100 and no striking difference in the behavior of the system was observed. Hence, we exclude finite size effects. The fact that already comparatively small lattices exhibit this independence from their size seems to indicate the existence of only weak spatial correlations between sites separated by a large distance. For high densities, i.e. close to or above the critical density $\eta_{cr}$, this statement does not remain valid. In fact, we expect the existence of long range correlations. However, in the present paper we are mostly concerned with the low density region and we will not devote ourselves extensively to the behavior of the automaton for high car density.

Each of the $L \times L$ automaton sites had its own four queues of length $Q = 10$ connecting the site with one of its four neighboring sites as described in the previous section. For each value of the density $\eta = \frac{N}{L \times L}$, where $N$ is the number of cars, we started the system at randomly chosen initial conditions. Each car was assigned the initial site, the queue in this site (i.e. whether it comes from North, East, South or West) and the desired direction (left, straight, right) for the next step. When a car transits to an empty site from a certain direction it occupies the first empty place in the queue. If it is at the top position of the queue it can transit to the next site or it has to wait, depending on the cars coming from the other directions of the same site, and depending on the traffic rules which apply to the given situation. If a car transits to another site and if there are other cars located in the previously mentioned queue, all of them advance by one position in this queue hence, moving up "closer to the crossing".

After starting the simulation for a given number of cars, i.e. for a given density $\eta$, we evolved the system 10,000 time steps without recording the statistics in order to let the system forget about the initial preparation. Typically a "stationary" behavior was found after only a few hundred time steps. In density regions very close to the transition value however, the relaxation of the system lasts longer which hints at some phase–transition–like phenomenon. The mean velocity of the cars was then computed using 2,000 time steps. We varied the car density beginning from $\eta = 0$ to $\eta = 3.8$ in steps $\Delta\eta = 0.01$.

The results of the simulations are shown in fig. 2. Like in simulations of cellular automata performed by other authors with different traffic rules we found a smoothly decaying function in the low density region. At a certain threshold value $\eta_{cr} \approx 1.6$ the character of the traffic flow changes abruptly and the system transits into the clogged regime. But different from other systems we find a remnant velocity in the clogged regime. This is due to the rule that in case all four of the queues are occupied there will be one car selected randomly which then will be allowed to move. Therefore, in our model there exist situations where a jamming at a crossing with four queues occupied may still dissolve.

The curve in the small density region is very smooth, i.e. there are almost no fluctuations. Hence, we can assume that long range correlations do not play a major role there. In the following section IV we will present a statistical description appropriate to model the flow in the low density regime, i.e. for $\eta \ll \eta_{cr}$, which was inspired by this observation.

## IV. A STATISTICAL DESCRIPTION OF VEHICULAR TRAFFIC

Let the number of cars be denoted by $N$ and the number of cells respectively crossings building the torus by $M = L \times L$ where $L$ is the common perimeter of the torus. For the global density we choose the symbol $\eta = N/M$.

The quantity we focus our attention on is a time averaged[2] and normalized velocity

$$\overline{v} = \frac{1}{T} \sum_{t=1}^{T} \frac{N - \Delta(t)}{N} = 1 - \frac{\langle \Delta \rangle_T}{N} \ , \qquad (1)$$

where $\Delta(t)$ means the total number of cars prevented from moving by traffic rules at time $t$ which depends on the exact situation experienced during the simulation.

For a description of the dynamics in the low density region $\eta \ll 1$ we apply the following arguments:

1. An effective description is possible which separates the process of clustering at the $M$ possible crossings from the process of obeying the traffic rules. Due to this indepen-dence assumption the time average (1) extracted from a simulation can be decomposed into two averages: first a time and cluster average related to the traffic rules and second a time average with respect to the clustering process.

---

[2]When performing the simulation the parameter $T$ is chosen sufficiently large in order to scan the state space respectively the limit set with sufficient accuracy.

2. Given $i$ cars meeting at one crossing a variety of situations is possible. Every car is approaching the crossing from either North, East, South or West and intends to go straight, to turn left or to turn right. In the long run all possible situations will have equal probability. Hence, the average number of cars which must wait, denoted by $\overline{\delta_i}$ with $0 \leq \overline{\delta_i} \leq i$, can be computed by combinatorial reasoning. The result of this computation is collected in table II for $i = 1, \ldots, 8$. A sketch of the derivation is given in appendix A.

   For clusters of size $i > 8$ the $\overline{\delta_i}$ are not shown since their contributions are suppressed by very small cluster probabilities.

3. Central to our description is the identification of cluster distributions, i.e. we write $\underline{k} = (k_0, k_1, \ldots, k_N)$ meaning that $k_i$ denotes the number of cells in the system where $i$ cars are meeting. Of course, there are two boundary conditions, namely

$$\sum_{i=0}^{N} k_i = M \tag{2}$$

and

$$\sum_{i=0}^{N} i \, k_i = \sum_{i=1}^{N} i \, k_i = N \ . \tag{3}$$

We now assume that for the low density regime $\eta \ll 1$ the process of clustering is ergodic which means for almost all initial configurations the time average can equivalently be represented by an average according to an invariant probability distribution. Note that this probability distribution is defined over the space of all configurations $\underline{k}$ compatible with the boundary conditions (2) and (3). In order to derive this distribution we additionally assume that the dynamics effectively is equivalent to a process of independently distributing $N$ cars (balls) among $M$ cells with crossings (urns) according to the equidistribution. This means we assume the discrete dynamics to be a Bernoulli process. Then the probability distribution reads

$$p(\underline{k}) \;=\; \frac{M!\,N!}{M^N \prod_{j=0}^{N} [(j!)^{k_j}\,k_j!]}\;. \qquad (4)$$

To understand formula (4) in detail one has to realize that there exist $M!/[k_0!k_1!\ldots k_N!]$ different ways to index the cells with numbers $0,1,\ldots,N$ since all cells containing the same number $i$ of cars are not distinguishable. Then there exist $N!/\left[(0!)^{k_0}(1!)^{k_1}\ldots(N!)^{k_N}\right]$ ways to index the cars with numbers $1,\ldots,N$ taking care of the fact that cars within an identical cell are not distinguishable. Finally the equidistribution with respect to all configurations yields the factor $M^{-N}$. This result was obtained by von Mises (1939) [49].

Now the time average $\langle\Delta\rangle_T$ is replaced by the ensemble average

$$\langle\Delta\rangle_K \;=\; \sum_{\underline{k}}^{*} p(\underline{k}) \sum_{i=1}^{N} k_i \overline{\delta_i}\,\alpha_i\;. \qquad (5)$$

The sum with respect to $i$ reflects the average delay due to the traffic rules at the crossings and the sum with respect to all configurations obeying (2) and (3) – which is indicated by the symbol $*$ – accounts for the cluster statistics. The factor $\alpha_i$ accounts for the fact that the dynamics of the real process differs from a Bernoulli–process. In general there may exist spatial as well as temporal correlations. The spatial correlations become important for densely filled lattices. The dynamics which control the traffic within one cell is strongly influenced by the cars moving in neighboring cells. Finally the emergence of a traffic jam clearly expresses these correlations. On the contrary temporal correlations play the dominant role for very sparse system, i.e. in the low density region. Consecutive situations are not completely statistically independent. Let us briefly explain this effect considering only cluster of size $i=2$: When the cars are randomly distributed amongst the lattice sites there is a certain probability that two cars $p$ and $q$ occupy the same site and hence form a cluster of size $i = 2$. Assuming statistical independence this probability is time invariant but for the cellular automaton the situation is quite different.

If two cars meet at a site there exist two possibilities: either both cars continue, or one of them continues and the other one has to give way. There is no chance that both cars meet again one time step later. In the extreme case of only two cars, i.e. $N = 2$, the probability for them to meet by random distribution is thus effectively reduced by a factor of two in comparison with the automaton dynamics. Given only a small number $N$ of cars, i.e $\eta = N/M \ll 1$, and provided only two of them $p$ and $q$ meet occasionally one has to consider in the following time step a system where only $N - 2$ cars feel the interaction with (in our case: can meet occasionally) $N - 1$ other cars and 2 cars which can meet only $N - 2$ cars, i.e. $p$ is not allowed to meet $q$ and $q$ cannot meet $p$. Hence, the system has some kind of memory and therefore it is not ideally Bernoullian.

Obviously this effect plays a crucial role only for small $N$ since otherwise the relative difference between $N - 1$ or $N - 2$ is negligible. Speaking in terms of density this means that the effect is substantial only for very small densities. Then however, clusters larger than $i = 3$ are suppressed by small likelihood. The net effect of these correlations can be taken into account through counting all possible configurations which can be realized after two cars have met. This procedure is similar to the method employed for the determination of the $\overline{\delta}_i$ (as explained in appendix A). Neglecting contributions related to clusters of size larger than three, which is safe for very small densities (see fig 3, explanation below) means $\alpha_i = 1$ for $i \geq 4$. From simple geometric considerations we find $\alpha_1 = 1$, $\alpha_2 = 1/2$ and $\alpha_3 = 7/4$.

Inserting expression eq.(5) into (1) yields

$$\overline{v} = 1 - \frac{\langle \Delta \rangle_K}{N} = 1 - \sum_{i=1}^{N} \frac{\overline{\delta}_i \alpha_i}{N} \sum_{k_1,\ldots,k_N}^{*} k_i \, p(k_1, \ldots, k_N) \tag{6}$$

and performing the sum over all $k_j$ ($j \neq i$) we find

$$\overline{v} = 1 - \sum_{i=1}^{N} \frac{\overline{\delta}_i \alpha_i}{N} \sum_{k_i=0}^{\lfloor N/i \rfloor} k_i \, p(k_i)$$

$$= 1 - \sum_{i=1}^{N} \frac{\overline{\delta_i} \alpha_i}{N} \langle K_i \rangle . \tag{7}$$

The distribution $p(k_i)$ – declared for $i = 0, 1, \ldots, \lfloor N/i \rfloor$ – can be calculated using the inclusion–exclusion principle [50,51]. This derivation is performed in appendix B and the result reads

$$p(k_i) = \frac{N!}{M^N} \sum_{j=k_i}^{M} (-1)^{(j-k_i)} \binom{j}{k_i} \frac{(M-j)^{(N-ji)}}{(i!)^l \, (N-ji)!} . \tag{8}$$

We see from (7) that the ingredients we actually need are the first moments $\langle K_i \rangle$ of the cluster distribution. They can be derived calculating a generating function for the (descending factorial) moments, denoted $H_i^{(M)}(z, x)$ [50]. This calculation is performed in appendix C. The result reads

$$\langle K_i \rangle = \binom{N}{i} \frac{1}{M^{(i-1)}} \left(1 - \frac{1}{M}\right)^{(N-i)} . \tag{9}$$

In fig. 3 we have depicted the expectation value of the cluster distribution $\langle K_i \rangle$ (eq. (9)) divided by the system size $M = L \times L$. The solid lines correspond to the function given in eq. (9) and the points represent related data taken from our simulation on a $50 \times 50$ torus. The values for $i = 0$ give the probability for a site to be empty. For $i = 0$ and $i = 1$ the analytical and numerical results agree perfectly, for larger $i$ the discrepancies between the solid lines and points are a direct consequence of the deviations from the assumed independent behavior as explained above. Note that the probabilities in fig. 3 are plotted using a logarithmic scale. The observed discrepancies for clusters of size larger than 3 will hardly have any influence on the overall behavior of the automaton. For values $\eta > \eta_{cr}$ the simulation data illustrate a breakdown for the small occupation numbers (except $i = 0$) due to the emergence of jams which are clusters of high order. At the transition point the occupation rates abruptly drop by three orders of magnitude. The precise position of this transition point is masked by the finite size of $N$ used in the simulation. Increasing the car density $\eta$ beyond the blurred critical zone results in a slow rise of the occupation number for the small clusters and simultaneously, in a slow decline of the number of empty sites. This is due to the increasing number of cars still moving around the jammed crossings.

Inserting equation (9) into (7) yields the following formula

$$\overline{v} = 1 - \sum_{i=1}^{N} \frac{\overline{\delta_i}\alpha_i}{N} \binom{N}{i} \frac{1}{M^{(i-1)}} \left(1 - \frac{1}{M}\right)^{(N-i)}. \tag{10}$$

The last step will be to substitute in (10) the density $\eta$ for the number of cars according to $N = \eta M$ which results in

$$\overline{v} = 1 - \sum_{i=1}^{\eta M} \frac{\overline{\delta_i}\alpha_i}{i!} \left(\eta - \frac{1}{M}\right) \cdots$$
$$\cdot \left(\eta - \frac{i-1}{M}\right) \left(1 - \frac{1}{M}\right)^{(\eta M - i)} \tag{11}$$

The main result of our statistical description, namely the mean velocity as a function of the car density (eq. (11)), is plotted in fig. 4. The solid line shows the function given by eq. (11) and the points are the data taken from the simulation on the $50 \times 50$ torus (compare fig.2. In the low density region we find a nice agreement. The closer one approaches the transition point the more eq. (11) overestimates the simulation results. This can be understood by recalling the fact that for densities close to the critical value the effective decomposition of the automaton dynamics into the cluster dynamics and the single site dynamics becomes inappropriate due to formation and dissolution of short lived jams (like critical fluctuations). They tend to increase the occupation number of larger clusters and hence, lead to an effective decrease of the average velocity.

In the thermodynamic limit – which means $M, N \to \infty$ and keeping $\eta = N/M$ constant – we drop terms of order $\mathcal{O}(M^{-1})$ and end up with

$$\overline{v} \stackrel{M \to \infty}{\approx} 1 - \sum_{i=1}^{\eta M} \frac{\overline{\delta_i}\alpha_i}{i!} \eta^{(i-1)} \exp(-\eta) \tag{12}$$

where the exponential corresponds to the dominant contribution of the last factor in (11). This formula can be expressed by a series expansion in powers of $\eta$

$$\overline{v} \stackrel{M \to \infty}{\approx} 1 - \sum_{i=1}^{\eta M} \sum_{k=0}^{\infty} \frac{(-1)^k \overline{\delta_i}\alpha_i}{i!k!} \eta^{(i-1+k)} \tag{13}$$

$$= 1 - \left(\frac{\overline{\delta_2}\alpha_2}{2}\right)\eta + \left(\frac{\overline{\delta_2}\alpha_2}{2} - \frac{\overline{\delta_3}\alpha_3}{6}\right)\eta^2$$
$$- \left(\frac{\overline{\delta_2}\alpha_2}{4} - \frac{\overline{\delta_3}\alpha_3}{6} + \frac{\overline{\delta_4}\alpha_4}{24}\right)\eta^3 + - \ldots$$

Note that there is no constant term in the double sum since $\overline{\delta_1} = 0$. Equation (13) can be truncated at a given order of $\eta$ hence, giving rise to an approximation scheme. In fig. 5 we plotted the linear (dashed), the quadratic (dotted) and cubic (long dashed) approximations together with the full formula (fat solid). The range of reliability visibly increases with increasing the order of approximation. On the other hand this scheme clearly illustrates that the functional relation between mean velocity and global density definitely is *nonlinear*, except for very small car densities.

## V. CONCLUSION

We investigated the behavior of a two–dimensional cellular automaton with periodic boundary conditions to simulate traffic flow in cities. The automaton mimics realistic traffic rules which apply to our everyday experience of vehicular traffic. In agreement with similar models we found a slow decay for the mean velocity $\langle v \rangle$ as a function of the global traffic density $\eta$. At a critical threshold value $\eta_{cr}$ the mean velocity collapses abruptly and the system transits into another regime of global behavior which we call the jamming regime. Beyond the critical density $\eta_{cr}$ the average velocity is very small and it declines further when increasing the density. By simulating different sizes of automata we could exclude the influence of finite size effect provided the density value is not rather close to the critical density.

Applying combinatorics and statistical methods for the description of the system in the low density regime the analytical calculation performed in section IV yielded the average car velocity as a function of the car density – $\langle \overline{v} \rangle (\eta)$ – which nicely agrees with the values of the numerical simulations in section III. Since the derived functional relationship between mean velocity and global traffic density was based on very general assumptions (conservation of the number of cars, weak spatial correlations in the low density regime, ergodicity) we expect this description to be valid for a wider class of traffic systems. Note that the specific structure of traffic rules enter the description only through the average delays $\delta_i$. The only nonlocal ingredient arises from the restriction that cars have to stop in case the next desired queue at a neighboring site is totally filled up. But this nonlocal character only becomes substantial for densities close to the critical value where it causes long range spatial correlations.

Several authors (e.g. [52]) assert that the average velocity $\overline{v}$ in the low density regime is a linear function of the density $\eta$ and indeed the simulation results seem to support this observation. A more detailed analysis however reveals that this function definitely is nonlinear. An analytic expansion shows (eq. (13), fig. 5) that the results become dramatically wrong when truncating the formula after the linear term. Surprisingly the analytic description gives good results even for densities not too small where the nonlocal effects of the dynamics and hence, long range correlations spoil the basic assumptions of our treatment.

For still higher values of the density $\eta$ the results of our simulation agree with results reported in literature (e.g. [52] and many others), but no longer with our approximate theory. Above a critical density we observe an abrupt transition of the system into the jammed regime where the averaged velocity is close to zero. This regime however is beyond the scope of our statistical description and has to be investigated starting from other approaches (e.g. nucleation processes).

**ACKNOWLEDGMENTS**

We thank H. Herzel and L. Schimansky–Geier for helpful discussion. We greatly acknowledge U. Küchler for drawing our attention to ref. [50] and for discussing the combinatorics in detail.

**APPENDIX A: CALCULATION OF THE AVERAGE NUMBER OF WAITING CARS $\overline{\delta_I}$ FOR CLUSTERS OF SIZE $I = 1$ AND $I = 2$**

In table II we present the average number $\delta_i$ of cars which are prevented from moving given $i$ cars meeting at a crossing. Obviously the number $\delta_i$ is confined to the interval $[0, i]$. The average is performed with respect to all possible configurations which are assumed to possess equal probability. The case $i = 1$ is rather simple: If there is only one car at the crossing it has never to give way hence, $\overline{\delta_1} = 0$. The case $i = 2$ is not that trivial and the computation of $\overline{\delta_2}$ requires some combinatorial effort. Since each of the cars can wait at one of the four sides of the crossing and can proceed in one of the three directions (left, straight on, right) we have $4 \times 3 \times 4 \times 3 = 144$ different situations of equal probability. Each of these situation can be assigned to one of the situations in the left column of table I. We now explain the columns of table I:

1. A sketch of the configuration given by a related graph.

2. The number of the different realizations that relate to the graph in the first column assuming first that the cars are distinguishable. The first factor $A$ in the form $A \times B \times C$ originates from the symmetry according to rotation by $\frac{1}{2}\pi$, $\pi$ and $\frac{3}{2}\pi$. The second factor $B$ denotes the number of choices for the first car (which is now assumed to come from South) and the last factor $C$ gives the number of situations possible for the second car.

3. When we take into account that the cars in fact cannot be distinguished the number of different events that belong to the figure in the first column has to be multiplied by the factor given in the third column.

4. The total number of events for the whole class denoted by the graph, i.e. (column 2× column 3).

5. The number of cars stopped by the traffic rules at this crossing in one of the possible realizations (either one or none).

6. The total number of cars which have to stop related to the whole class denoted by the graph, i.e. (column 4× column 5).

From the last line in table I one sees that there are 144 different situations of equal probability which amounts to 288 cars. Furthermore we see that 92 of those 288 cars have to stop. Hence, the average number of the cars stopped for 2–clusters is $\overline{\delta_2} = 92/144 \approx 0.639$.

For the clusters of size $i = 3$ the average number of cars which are prevented from moving $\overline{\delta_3}$ can be calculated in analogy however the corresponding table contains about 8 times as much situations as table I. Therefore we do not want to present it here. The result is $\overline{\delta_3} = 1.479$. For cluster of size larger than 3 the corresponding $\overline{\delta_i}$ have been calculated by the method of complete enumeration using a computer. These empirical results are collected in table II. For $i = 1 \ldots 3$ they are identical with the analytical results.

For completeness sake we remark that for large cluster sizes ($i > 8$) there exists a negligible probability that one of the directions of the crossing is not occupied. Typically there are many cars waiting in each of the four queues. In those very likely cases our traffic rules allow exactly one (randomly chosen) cars to drive, all other $i - 1$ cars have to stop. Therefore we find $\overline{\delta_i} \stackrel{i \gg 1}{\to} i - 1$. Certainly the contributions of such big clusters will only play a minor role in our calculations due to very small likelihood.

## APPENDIX B: THE DERIVATION OF THE CLUSTER PROBABILITIES

The probability $p(k_i)$ to find exactly $k$ sites which are occupied each by $i$ (independently moving) cars in a system of $M$ crossings and $N$ cars is given in eq. (8). The derivation of this formula employing the inclusion–exclusion principle will be performed in this appendix. Note that the problem to find $p(k_i)$ is different from the trivial problem to find the probability for the $k_i$ crossings which are occupied by *at least i* cars.

The inclusion–exclusion principle relates the probabilities for a finite number of sections of events to the probabilities of an exact number of events.

Let $(\Omega, P)$ be a probability space, $A_1, \ldots, A_N$ be events, and for arbitrary $\{m_1, \ldots, m_j\} \subset \{1, \ldots, N\}$ let $P(A_{m_1} \cap \ldots \cap A_{m_j})$ be known probabilities. We define

$$S_j = \sum_{m_1, \ldots, m_j} P(A_{m_1} \cap \ldots \cap A_{m_j}) \tag{B1}$$

Moreover, let be $B_n = \{\omega \in \Omega : \omega \in A_m \text{ for exactly } n \text{ values of } k\}$. Then the inclusion–exclusion principle asserts that

$$P(B_n) = \sum_{j=n}^{N} (-1)^{(j-n)} \binom{j}{n} S_j \qquad (B2)$$

Applied to our case the $S_j$ are the probabilities that $j$ cells each contain $i$ cars and the rest of the cars, i.e. $(N - ji)$ cars, are distributed arbitrarily among the remaining $(M - j)$ cells; this probability can be derived with ease and reads

$$S_j = \frac{N!}{(i!)^j (N - ji)!} \left(\frac{1}{M}\right)^{(ji)} \left(1 - \frac{j}{M}\right)^{(N-ji)} \qquad (B3)$$

To explain this formula we first index each of the $N$ cars with the numbers $1, \ldots, N$ which yields the factor $N!$. Since the $i$ cars within each of the $j$ cells are not distinguishable we have to divide this factor by $(i!)^j$ and since the remaining $(N - ji)$ cars are not distinguishable too additionally by $(N - ji)!$. After having numbered the cars we subsequently fill the first cell with cars $1, \ldots, i$, the second cell with cars $(i + 1), \ldots, (2i)$ and so on. In this way we distribute the cars numbers $1, \ldots, ji$ among the urns $1, \ldots, j$ yielding the factor $M^{(-ji)}$. The remaining $(N - ji)$ cars are distributed successively among the remaining $M - j$ cells at random which explains the factor $[(M - j)/M]^{(N-ji)}$.

Insertion of this probability into the inclusion–exclusion formula (B2) yields

$$\begin{aligned} p(k_i) &= \sum_{j=k_i}^{M} (-1)^{(j-k_i)} \binom{j}{k_i} S_j \\ &= \frac{N!}{M^N} \sum_{j=k_i}^{M} (-1)^{(j-k_i)} \binom{j}{k_i} \frac{(M-j)^{(N-ji)}}{(i!)^j (N - ji)!} \end{aligned} \qquad (B4)$$

Because of the generalized definition of factorials – using the gamma function – the sum effectively only ranges from $k_i$ up to $\lfloor M/i \rfloor$ which is sensible.

# APPENDIX C: CALCULATION OF THE FIRST MOMENTS OF THE CLUSTER DISTRIBUTION $\langle K_I \rangle$

In the following we will derive the mean value of the cluster distribution $\langle K_i \rangle$ which is given in eq. (9).

The generating function for the (descending) factorial moments of the cluster distribution we use is

$$H_i^{(M)}(z,x) = \sum_{N=0}^{\infty} \sum_{k_i=0}^{\infty} \frac{M^N z^N}{N!} x^{k_i} p(k_i, N) \tag{C1}$$

where $p(k_i, N)$ is the probability to find a value $k_i$ for the stochastic variable $K_i$ when trying with $N$ balls. The benefit of such a complicated looking generating function is that the sums can be performed yielding an analytical expression namely

$$H_i^{(M)}(z,x) = \left[\exp^z + \frac{z^i}{i!}(x-1)\right]^M \tag{C2}$$

For the explicit derivation the reader is referred to the book by Johnson and Kotz[3] p. 116ff [50]. Moreover, this function is related to the (descending) factorial moments of the stochastic variable $K_i$ according to

$$\frac{N!}{M^N} \frac{d^r}{dx^r} \left(H_i^{(M)}(x,z)\right)\bigg|_{x=1}$$
$$= \sum_{N=0}^{\infty} z^N \sum_{k_i=0}^{\infty} k_i(k_i-1)\ldots(k_i-r+1) p(k_i, N)$$
$$= \sum_{N=0}^{\infty} z^N \left\langle k_i(k_i-1)\ldots(k_i-r+1) \right\rangle \tag{C3}$$

hence, the first moment $\langle K_i \rangle$ is given as the coefficient of $z^N$ in

$$\frac{N!}{M^N} \frac{d}{dx} \left(H_i^{(M)}(z,x)\right)\bigg|_{x=1} \tag{C4}$$

Consequently we calculate

$$\frac{N!}{M^N} \frac{d}{dx} \left(\left[\exp^z + \frac{z^i}{i!}(x-1)\right]^M\right)\bigg|_{x=1}$$
$$= \frac{N!}{M^{(N-1)}} \exp^{[z(M-1)]} \frac{z^i}{i!}$$
$$= \sum_{k=0}^{\infty} \frac{N!}{i!\, k!} \frac{(M-1)^k}{M^{(N-1)}} z^{(k+i)} \tag{C5}$$

---

[3]one has to identify $j \equiv i, m \equiv M, n, \equiv N, M_i \equiv K_i, g \equiv k_i, \Pr[M_j = g|n] \equiv p(k_i)$

and finally arrive at

$$\begin{aligned}\langle K_i \rangle &= \frac{N!}{i!\,(N-i)!}\,\frac{(M-1)^{(N-i)}}{M^{(N-1)}} \\ &= \binom{N}{i}\,\frac{1}{M^{(i-1)}}\,\left(1-\frac{1}{M}\right)^{(N-i)} \end{aligned} \qquad (C6)$$

TABLES

TABLE I. All situations which might occur when two cars meet at a crossing, their frequency, their symmetry according to rotation and due to indistinguishability of the cars and the number of cars which have to stop in the current situation (explanation in the text).

| graph | rot. symm. | dist. | #events | waiting | #stopping |
|---|---|---|---|---|---|
| 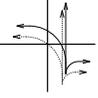 | $4 \times 3 \times 3$ | 1 | 36 | 1 | 36 |
| 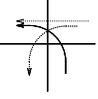 | $4 \times 1 \times 2$ | 2 | 16 | 1 | 16 |
| 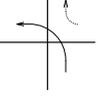 | $4 \times 1 \times 1$ | 2 | 8 | 0 | 0 |
| 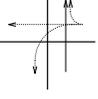 | $4 \times 1 \times 3$ | 2 | 24 | 1 | 24 |
| 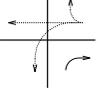 | $4 \times 1 \times 3$ | 2 | 24 | 0 | 0 |
| 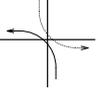 | $2 \times 1 \times 1$ | 2 | 4 | 0 | 0 |
| 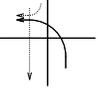 | $4 \times 1 \times 2$ | 2 | 16 | 1 | 16 |
| 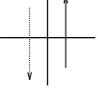 | $2 \times 1 \times 1$ | 2 | 4 | 0 | 0 |
| 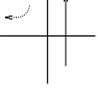 | $4 \times 1 \times 1$ | 2 | 8 | 0 | 0 |
| 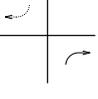 | $2 \times 1 \times 1$ | 2 | 4 | 0 | 0 |
| | total | | 144 | | 92 |

TABLE II. The average number of cars which are stopped $\overline{\delta_i}$ when $i$ cars meet at a crossing.

| #cars meeting $i$ | $\overline{\delta_i}$ |
|---|---|
| 1 | 0.0 |
| 2 | 0.638889 |
| 3 | 1.479167 |
| 4 | 2.467014 |
| 5 | 3.524740 |
| 6 | 4.604709 |
| 7 | 5.683838 |
| 8 | 6.752964 |

FIGURES

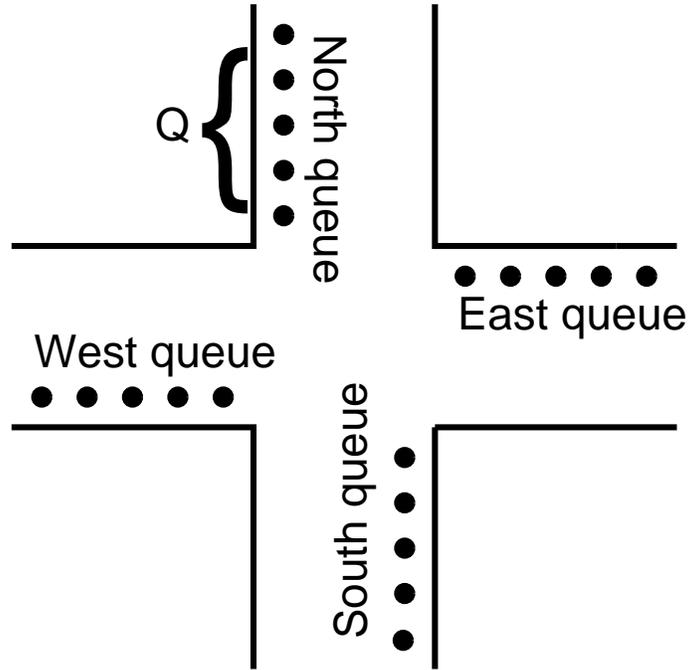

FIG. 1. The schematic plot of a crossing

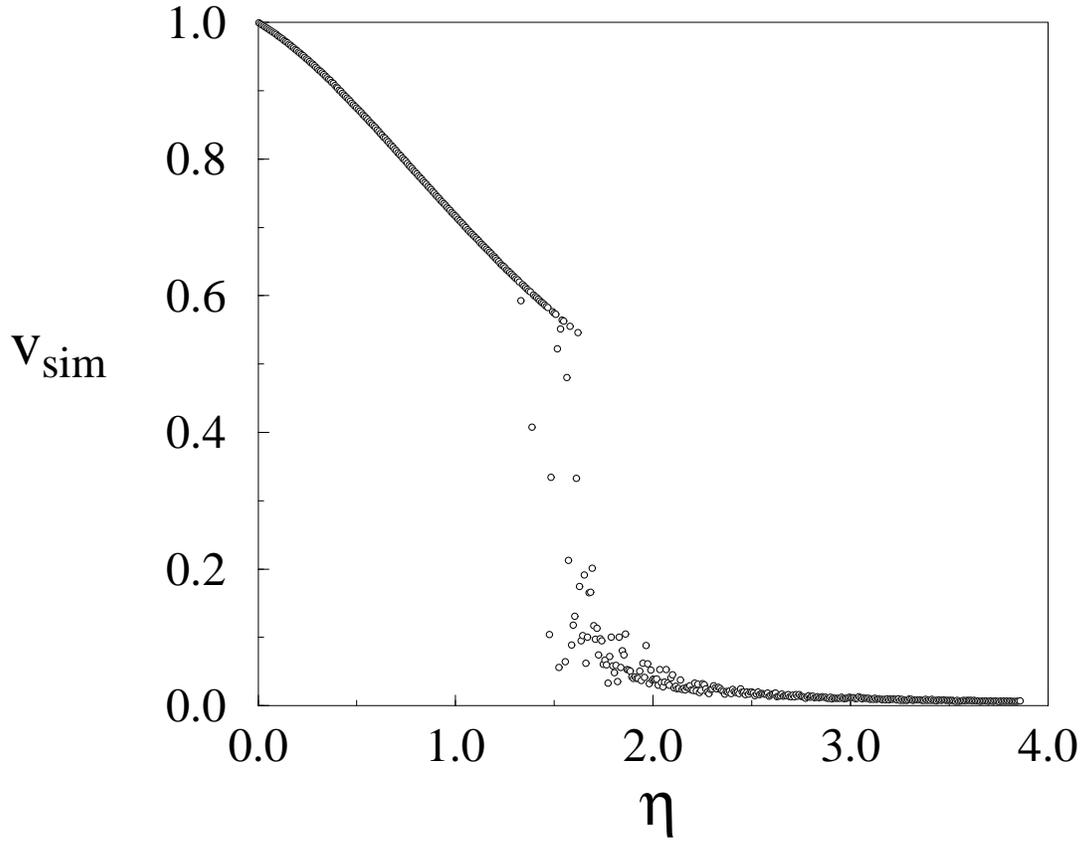

FIG. 2. The simulation result: The mean velocity $v_{sim}$ of $N$ cars on a periodic lattice of size $L \times L$ vs the global traffic density $\eta = N/(L \times L)$. Except for values around the critical density $\eta_{cr}$ the curve is rather smooth. In the vicinity of the transition zone critical fluctuations, i.e. temporary jams, cause an irregular relationship. This can be understood from the fact that the typical lifetime of jams becomes of the same order as the simulation time.

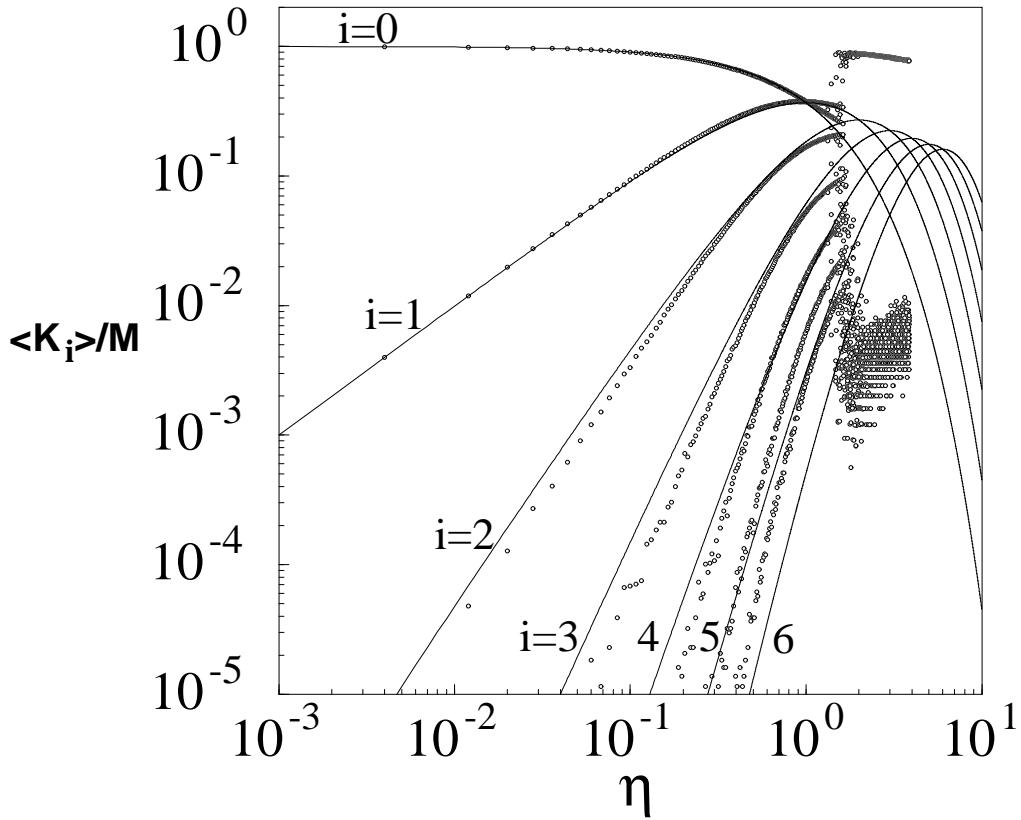

FIG. 3. The average number of clusters of size $i$ normalized to the lattice size $M = L \times L$ as a function of the global traffic density $\eta = N/M$. The lines represent eq. (9) while the points correspond to numerical data achieved by a simulation on a $50 \times 50$ lattice (torus). The curve $i = 0$ gives the probability for a site to be empty.

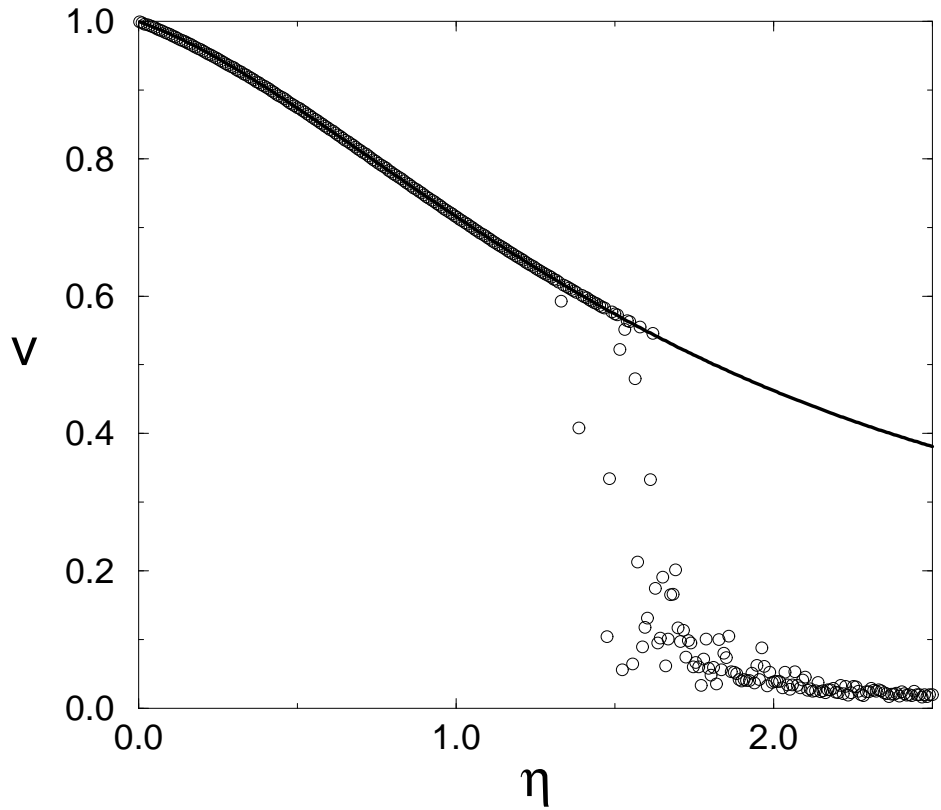

FIG. 4. The average velocity vs the global traffic density for the low density regime (non jamming traffic). The curve displays the result from equation (11), the points show data taken from the simulation.

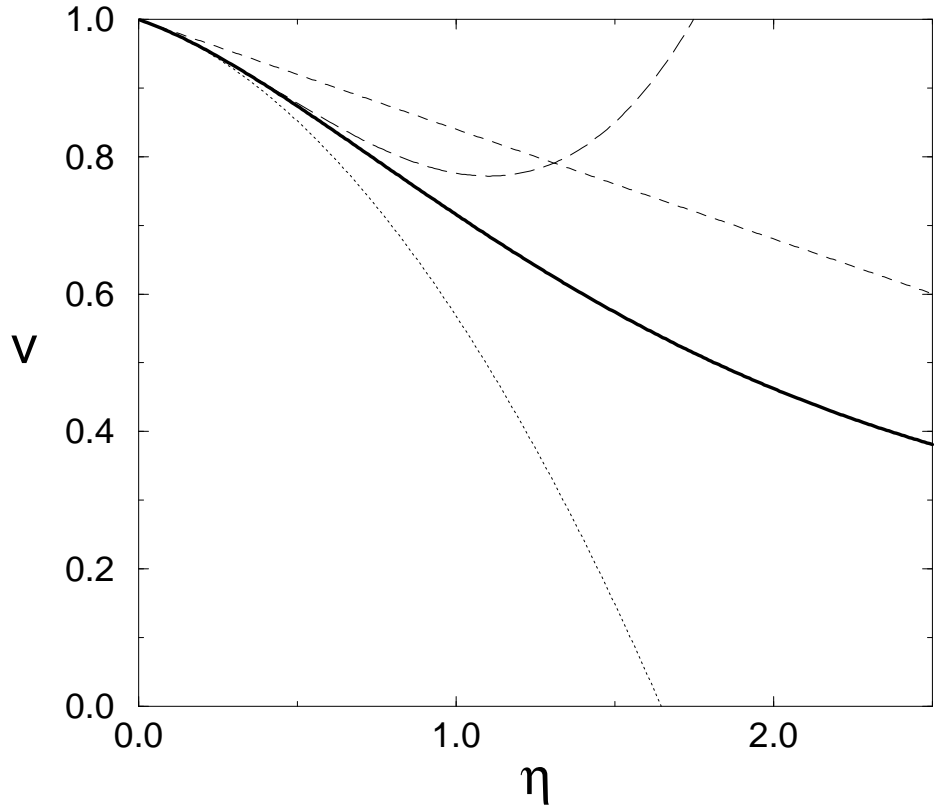

FIG. 5. The average velocity as derived from the statistical description (eq. (9)) (fat line) and truncated series expansions: linear (dashed), quadratic (dotted) and cubic (long dashed). This plot clearly illustrates a nonlinear relationship between average velocity and global density.